\documentclass[epj]{svjour}
\usepackage{graphics}
\begin{document}
\title{Evolution of single-particle structure of silicon isotopes }
\author{O.V.~Bespalova\inst{1}\thanks{\email{besp@sinp.msu.ru}},  N.A.~Fedorov\inst{2}, A.A.~Klimochkina\inst{1}, M.L.~Markova\inst{2}, T.I.~Spasskaya\inst{1} \and T.Yu.~Tretyakova\inst{1}\thanks{\email{tretyakova@sinp.msu.ru}}
}                     
%
%
\institute{Skobeltsyn Institute of Nuclear Physics, Lomonosov Moscow State University, \\119991, GSP-1, Leninskie gory 1(2), Moscow, Russia \and Faculty of Physics, Lomonosov Moscow State University, 119991, GSP-1, Leninskie gory 1(2), Moscow, Russia}
\date{Received: date / Revised version: date}
%
\authorrunning{O.V. Bespalova et al.}
\titlerunning{ Evolution of single particle structure of silicon isotopes}

\abstract{
New data on proton and neutron single-particle energies $E_{nlj}$  of Si isotopes with neutron number $N$ from 12 to 28 as well as occupation probabilities $N_{nlj}$ of single particle states of stable isotopes $^{28,30}$Si near the Fermi energy were obtained by the joint evaluation of the stripping and pick-up reaction data and excited state decay schemes of neighboring nuclei. The evaluated data indicate following features of single-particle structure evolution: persistence of  $Z = 14$ subshell closure with $N$ increase, the new magicity of the number $N = 16$, and the conservation of the magic properties of the number $N = 20$ in Si isotopic chain. The features were described by the dispersive optical model. The calculation also predicts the weakening of $N = 28$ shell closure and demonstrates evolution of bubble-like structure of  the proton density distributions in neutron-rich Si isotopes.
} 
\maketitle
\section{Introduction}
\label{intro}

\begin{figure}
\resizebox{8,5 cm}{!}{%
\includegraphics{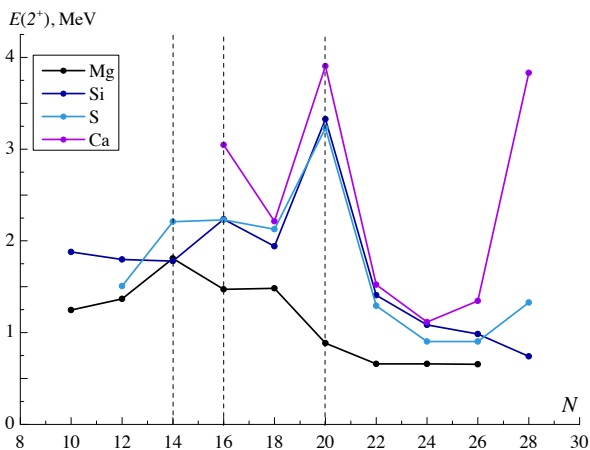}
}
\caption{(Colour online) Evolution of first excited states energies $E(2^+_1)$ in Ca, S, Si and Mg isotopic chains. Data from \cite{NNDC}.}
\label{fig:E2}      
\end{figure}

One of the main goals of modern nuclear physics is to study the evolution of nucleus structures of nuclei while shifting to exotic ones up to the proton and neutron drip lines. Significant progress of experimental techniques  gave a clue  for the proper understanding of the evolution and allowed to discover specific features of nuclei, appearing in the region far from $\beta$-stability line. Local appearance of new and disappearance of classic magic numbers of nucleons are among them. Such numbers as 8, 20, 28 tend to be found locally weakening throughout isotopic and isotonic chains. Otherwise, some signs of magic features can be traced for the nuclei with neutron and proton numbers $N, Z=6,16,32$, etc \cite{Sorlin,Ga15}. For example, the tendency of $E(2^+_1)$ to decrease for $^{42}$Si and $^{44}$S in comparison with doubly magic nucleus $^{48}$Ca can point to $N=28$ shell closure weakening in these isotopes (see Fig.~\ref{fig:E2}). Indeed, in \cite{Bastin,Tarpanov} $N=28$ in the isotone chain from $^{48}$Ca to $^{42}$Si was proved to demonstrate a gradual decrease of nuclear rigidity and the concomitant loss of its magic properties. At the same time, magic number $N=20$ persists in the chain from $^{40}$Ca to $^{34}$Si with no significant loss of its strength.

On the other hand the dependence of $E(2^+_1)$ from $N$ demonstrates a sort of local peak nearby $N, Z=16$ and $Z=14$ for the isotopes $^{30}$Si,$^{30}$S and $^{32}$S. These numbers reveal almost the same behavior as  magic ones. Both appearance and disappearance of magic nuclei are related to broadening and narrowing of energy gaps adjoining to the corresponding subshells. Such shifts are an inalienable part of general evolution. As it was already shown \cite{Ga15},  silicon isotopic chain is one of significant candidates for the detailed investigation in this case. Its stable isotopes, $^{28}$Si, $^{29}$Si and $^{30}$Si, may provide a bulk of experimental data of different reliability. So, an appropriate selection procedure and analysis can be applied for them. 
 Silicon isotopic chain allows to investigate the behavior of classic magic numbers $N=20$ and $N=28$ to prove the predictions mentioned before. Isotope $^{42}$Si  is an object of special interest, since it is exotic isotope for which one can investigate the influence of neutron excess on the features of $N=28$ shell \cite{Ga15,FW05}. The latest advances in the radioactive beam creation allow to follow neutron gathering from the lightest silicon isotopes up  to the  isotope $^{42}$Si \cite{Bastin,FW05,Fr06,Ta12}. 

The structural evolution can be revealed not only in the general picture of subshell shifts, but in particular tendencies for the proton and neutron densities. The  unusual behavior of proton density in $^{34}$Si was predicted a long time ago and used to be discussed in plenty of papers, being proved by different models (see, for example \cite{ML17,GG09}). The calculated proton density presents a bubble-like structure with a significant depletion in a central region. But whether it remains or changes significantly in the silicon isotope chain is a relevant question for analysis. 

The method of joint evaluation of the stripping and pickup reaction data on the same nucleus is applied in the present paper to obtain new evaluated data on evolution of single-particle energies and occupation probabilities of neutron and proton states in Si isotopes. Analysis of experimental information is complicated by the fact that stable silicon isotopes belong to the region of nuclei with stable deformation. The experimental value of quadrupole deformation of $^{28}$Si is $\beta_2 = -0.42 \pm 0.02$ \cite{HL84}. The magnitude of deformation of the remaining isotopes has not yet been determined exactly and, since different theoretical approaches lead to different results, requires a more accurate experimental study. In this situation, with insufficient experimental information, the spherical approximation can be justified as an initial estimate. It should also be noted good results were obtained for $^{28}$Si in dispersive optical model (DOM) in spherical approach \cite{AO12}. 
 
 At the end of the last century, the semiphenomenological dispersive optical model (DOM) was developed by C. Mahaux, R. Sartor, and their coauthors (see \cite{MS91} and references therein) by an example of magic nuclei $^{40}$Ca, $^{90}$Zr, $^{208}$Pb. Then, DOM was successfully applied to various nuclei, for example $^{86}$Kr\cite{Joh89}, $^{120}$Sn\cite{Zem04}, $^{51}$V\cite{Law89}. Further, the dispersive coupled-channels optical model potential was derived and applied to calculate quasielastic $^{232}$Th$(p,n)$ and $^{238}$U$(p,n)$  scattering cross sections \cite{Que07}. The success of the model made it possible to determine the global parameters of the dispersive optical model potential \cite{Mor07,Bes07} and to extrapolate the potential for some isotopic chains far from beta-stability line \cite{CM07,Bes15}. The most consistent application of the dispersion approach - dispersive self-energy method,  was developed and implemented in the Green's function method of the Dyson equation \cite{Dick17}. The version of DOM \cite{MS91} is applied in the present paper to calculate single-particle structure evolution of Si isotopes. In the second section the joint evaluation procedure of pickup and stripping reactions data is presented.  The dispersive optical model  used in this work  is considered in the third section. In the fourth section the results on the single particle state evolution and proton densities are discussed.

\section{Evaluation based on experimental spectroscopy}
\label{sec:2}
\subsection{Joint analysis of single-nucleon transfer reactions}
\label{sec:2:1}
One of the most informative and reliable experimental sources for the evaluation of nuclear single-particle states is presented by single-nucleon transfer reactions \cite{Bohr}. Previously, a lot of work was done to compile and evaluate the experimental data \cite{Endt}, but since the release of these works a significant amount of new data has appeared. Despite this, transfer reactions are always related to the experimental uncertainties (in angular momentum and related cross-section determination), which is why modern data require appropriate selection procedure as well.

In the framework of DWBA calculations for pickup and stripping reactions it is possible to define stripping $(+)$ and pickup $(-)$ strengths of $i$-states $G^{\pm}_{nlj}$ as the proportionality coefficient among the experimental  $\sigma_{exp}$ and the theoretical $\sigma_{DWBA}$ cross-sections:

\begin{equation}
\frac{d\sigma}{d\Omega}(l,j,\theta)_{exp}=W_{\sigma}G_{nlj}^{\pm}\frac{d\sigma}{d\Omega}(l,j,\theta)_{DWBA},
\label{dbwa}
\end{equation}

\noindent 
where $W_{\sigma}$ is a theoretical normalization factor. 

The correct normalization of spectroscopic strengths is a basic experimental problem for further precise calculations. Both this normalization and evaluation of single-particle energies are based on the joint numerical analysis of two complementary data sets of stripping and pickup experiments on the same target \cite{BV89,Pf}. This approach includes renormalization of spectroscopic strengths based on the sum rule for each subshell in the immediate neighbourhood of the Fermi surface:

\begin{equation}
f^+G_{nlj}^{+}+f^-G_{nlj}^{-}=2j+1 ,
\label{fsum}
\end{equation}

\noindent where $f^{\pm}$ are the normalizing coefficients. The spectroscopic strength $G_{nlj}$ of a group of levels connects the spectroscopic factors with occupation probabilities by standard sum rules \cite{FMc61}. For stripping reaction $$G_{nlj}^+ = \sum_i G_{nlj}^+ (i)= \sum_i\frac{2J_f+1}{2J+1}C^2S^+_{nlj}(i),$$
where $S^+_{nlj}$ denotes the spectroscopic factor, $J$ and $J_f$ refer to the spins of the target and final states, respectively, and $C$ is the isospin coupling factor. 
Applying different independent experimental data on pickup and stripping reactions on the same target with the sum rule consideration gives grounds for assuming sufficient reliability of the results obtained.

Analysis of spectroscopic strengths $G_{nlj}^{\pm}(i)$ would assist to define the $nlj$ level energy disposition \cite{MS93}:
\begin{equation}
E_{nlj}=\frac{G_{nlj}^{+}E^{+}_{nlj}+G_{nlj}^{-}E^{-}_{nlj}}{G_{nlj}^{+}+G_{nlj}^{-}},
\label{enlj}
\end{equation}
 
\noindent 
 where $E^{+}_{nlj}$ and $E^{-}_{nlj}$ are:
 \begin{eqnarray*}
 E^{+}_{nlj}&=&-S(A+1)+C^{+}_{nlj},\label{E+}\\ 
 E^{-}_{nlj}&=&-S(A)-C^{-}_{nlj}.\label{E-}
  \end{eqnarray*}
\noindent
Here  $S(A)$ is a proton or neutron separation energy for the $A$ nucleus (for proton and neutron subshells correspondingly), $C^{\pm}_{nlj}$ is the energy centroid as an average excitation energy for the $nlj$ set:

\begin{equation}
C_{nlj}^{\pm}=\frac{\sum_{i}E^{\pm}(i)G_{nlj}^{\pm}(i)}{\sum_{i}G_{nlj}^{\pm}(i)}.
\label{C}
\end{equation}

Subshell occupation probabilities $N_{nlj}$ are normalized on the full possible number of the nucleons $2j+1$ for the $nlj$ subshell and defined as \cite{MS93}:
\begin{equation}
N_{nlj}=\frac{G_{nlj}^{-}-G_{nlj}^{+}+2j+1}{2(2j+1)}.
\label{nnlj}
\end{equation}
These $N_{nlj}$ values might be approximated by the BCS-function in order to get estimations of the Fermi energy $E_{F}$ and the gap parameter $\Delta$:
\begin{equation}
N_{nlj}=\frac{1}{2}\left(1-\frac{E_{nlj}-E_{F}}{\sqrt{(E_{nlj}-E_{F})^{2}+\Delta^{2}}}\right).\label{BCS}
\end{equation}

The standard method  of centroids of single particle spectrum function, used in our work, strongly depends on the spectroscopic factor experimental values and on quality of visible spectrum. The single-particle energies extraction requires collecting the full fragmentation up to high energy \cite{DH12}. Moreover, the extraction of spectroscopic factors in experimental works depends on reaction model parameters and they differ in different experiments. The main purpose of the joint numerical analysis of one-nuclear transfer reaction is to compensate the inconsistency of the experimental data by selecting the most consistent results both stripping and pickup reactions with sufficient details and energy coverage.       

As long as even-even silicon isotopes under investigation ($A=26\div42$) include two stable nuclei  $^{28}$Si and $^{30}$Si, it would be convenient to combine the theoretical approach (see \ref{sec:2:2}, \ref{sec:2:3}) and the experimental data set for them. Moreover, according to the independent particle model (IPM) they should express closed $1d_{5/2}$ and $2s_{1/2}$ subshells correspondingly, so the following approach would reveal the best results mostly for the $1d2s$ shell and the adjacent shell $1f_{7/2}$.

Significant experimental data base for these nuclei has been gathered since the pickup and stripping reactions were first carried out. However, most of these experiments are dated 1960-1970 years, so they does not display  detailed spectroscopy and results does not cover a significant energy range. Despite this, variety of available data allows to put $f^+=f^-=1$ in (\ref{fsum}), omit the renormalization procedure  and apply the experimental spectroscopies directly to evaluate the single-particle energies. 16~experiments on the $^{28}$Si target and 19~experiments on the $^{30}$Si target were used in our analysis  in order to obtain single-particle energies and corresponding occupation probabilities of nucleon subshells in $^{28,30}$Si.

\begin{figure}
\resizebox{8 cm}{!}{%
\includegraphics {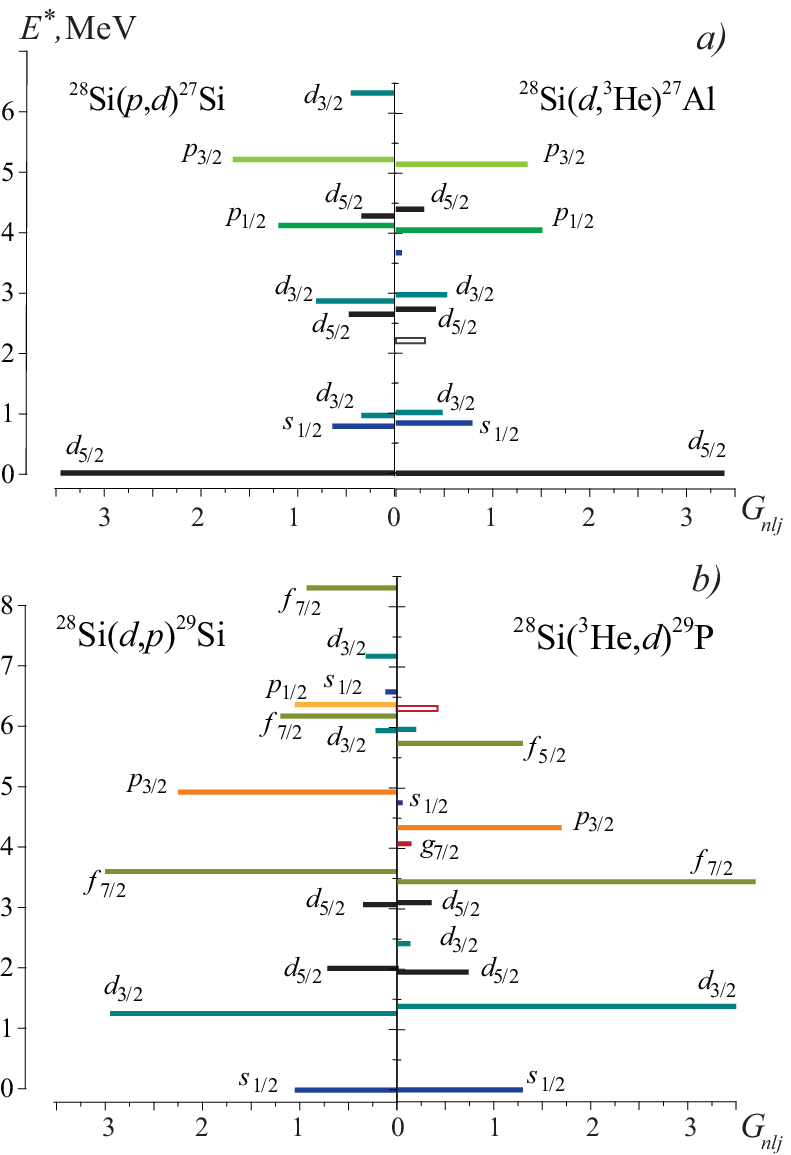}
}
\caption{(Colour online)  Spectroscopy of $^{28}$Si from one-nucleon transfer reaction, adopted in the joint numerical analysis.  The spectroscopic strengths $G_{nlj}$ for a) pickup of neutron \cite{Ko68} and proton \cite{Ma74} and b) stripping of neutron \cite{Me71} and proton \cite{Dy76}.}
\label{fig:sf28}      
\end{figure}

\begin{table}
\caption{Selected proton and neutron pickup and stripping reactions}
\label{tab:1}       
\begin{center}
\begin{tabular}{ccccc}
\hline\noalign{\smallskip}
&\multicolumn{2}{c}{\small{$^{28}$Si}} & \multicolumn{2}{c}{\small{$^{30}$Si}}\\
\noalign{\smallskip}\hline\noalign{\smallskip}
Reaction & $E$, MeV & Ref. & $E$, MeV & Ref. \\
\noalign{\smallskip}\hline\noalign{\smallskip}
($^3$He$,d$) & 25 & \cite{Dy76} & 25 & \cite{Ve90}\\

 ($d,^3$He) & 52 & \cite{Ma74} & 52 & \cite{Ma74}\\

 $(d,p)$ & 18 & \cite{Me71} &  12.3 & \cite{Pi00}\\

 $(p,d)$ & 33.6 & \cite{Ko68} &  27 & \cite{Ha75,Jo69}\\

\hline\noalign{\smallskip}
\end{tabular}
\end{center}
\end{table}

\begin{figure}
\resizebox{8 cm}{!}{%
\includegraphics {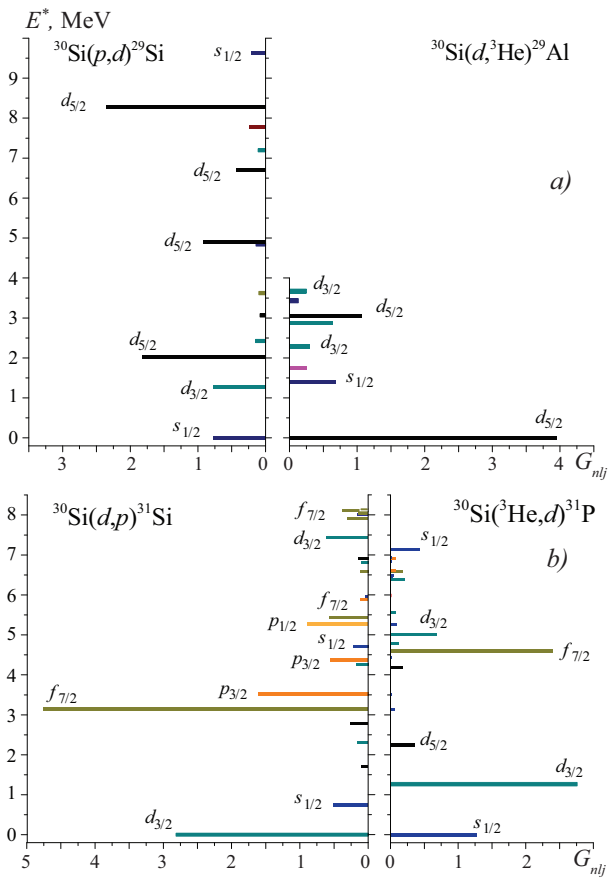}
}
\caption{(Colour online)  Spectroscopy of $^{30}$Si from one-nucleon transfer reaction, adopted in the joint numerical analysis.  The spectroscopic strengths $G_{nlj}$ for a) pickup of neutron \cite{Ha75} and proton \cite{Ma74} and b) stripping of neutron \cite{Pi00} and proton \cite{Ve90}.}
\label{fig:sf30}      
\end{figure}

The important part of the joint numerical analysis is to  select two complementary data sets of stripping and pickup experiments. The following set of selection criteria was used in the present work:
\begin{enumerate}
    \item Total pickup spectroscopic strength $G_{sd}^{-}$ value for the $1d2s$ shell should reveal the possible total number of protons or neutrons $N_p$ for the corresponding isotope (approximately $6$~protons and $6$~neutrons for $^{28}$Si, $6$~protons and $8$~neutrons for $^{30}$Si).
    \item Total stripping spectroscopic strength $G_{sd}^{+}$ value for the $1d2s$ shell should reveal the possible total number of nucleon holes $N_h$ for the corresponding isotope (approximately $6$~proton holes and $6$~neutron holes for $^{28}$Si, $6$~proton holes and $4$~neutron holes for $^{30}$Si).
    \item The current spectroscopic approach implies $$G_{nlj}^{+}+G_{nlj}^{-}=2j+1$$ for each $1d2s$ subshell, so it's required to minimize the absolute value of $a_{j}$ :
    \begin{equation}
    a_{j}=1-\frac{G^{+}_{nlj}+G^{-}_{nlj}}{2j+1}.
    \end{equation}\label{aj}
    \item The Fermi energy $E_{F}$ and $\Delta^2$ approximation (\ref{BCS}) errors $\sigma(\Delta^2)$ and $\sigma(E_{F})$ are taken into consideration.
    \item $E_F$ and $\Delta$ as the approximation parameters in (\ref{BCS}) should fit rough $E_{F}$ and $\Delta$ values:
  \begin{eqnarray}
  & E_{F}(A)=-\frac 12 (S(A)+S(A+1)),   \label{EF}\\
  & \Delta=-\frac{1}{4}(S(A+1)-2S(A)+S(A-1)),  \label{Dn}
     \end{eqnarray}
calculated from nuclear mass data \cite{AME12}.

    \item As long as $^{28}$Si represents equality for $N=14$ and ${Z=14}$ as the numbers of neutrons and protons, one may observe rough correspondence between proton and neutron pickup spectroscopy, proton and neutron stripping spectroscopy. For other isotopes correspondent isospin relations should be taken into account \cite{FMc61}.
\end{enumerate}

Since  neutron pickup spectroscopy reveals a sort of level limitation close to high energies, it might be valuable to supplement with a proton pickup spectroscopy using spectroscopic correspondence for isobaric analogue states of $^{29}$Si and $^{29}$Al with isospin $T=3/2$:
$$G_{n}^{-}(i)=\frac{G_{p}^{-}(i)}{2T+1}. $$

On the base of criteria (1)--(4)  the penalty function was generated:
\begin{eqnarray*}
p_k &=& \frac{1}{7}\left(\langle a_k \rangle + d^-_k + d^+_k + \frac{N^+_{max}-N^+_k}{N^+_{max}}+ \frac{N^-_{max}-N^-_k}{N^-_{max}} +\right.\\
 &&\left.+\frac{\sigma_k(E_F)}{\sigma_{max}(E_F)} + \frac{\sigma_k(\Delta^2)}{\sigma_{max}(\Delta^2)}\right),
\end{eqnarray*}
where $k$ indicates the pickup-stripping pair of experiments, $\langle a_k \rangle$ is the  value of $a_j$ (\ref{aj}) averaged over $j$-shells considered. The second and third terms optimize the particle and hole numbers on $sd$-shell in accordance with criteria (1) and (2):
$$ d^-=N_p-\sum_{sd} G^-_{nlj}, \,\,\,\, d^+=N_h-\sum_{sd} G^+_{nlj}.$$
Two next terms
take into account the deviation of the number of states in $k$-pair from the maximum value in stripping $(+)$ and pickup $(-)$ reactions in this sample of experiments thereby effectively taking into account the states fragmentation and energy coverage of data.  The last two terms are responsible for  the error minimization in determining $E_F$ and $\Delta^2$ by experimental data extrapolation. This condition imposes some requirement of correspondence of the BCS theory, which  seems appropriate in this case because of the strong configuration mixing in stable Si isotopes.
  
This penalty function was used to select proper experimental pairs with the maximum numbers of available spectroscopic states by force of $a_j$, $E_{F}$ and $\Delta$ errors minimization. Selected proton and neutron pickup and stripping reactions are presented in Tabl.~\ref{tab:1}. Spectroscopic strengths for single-particle reactions on $^{28}$Si show a very good correspondence between proton and neutron spectroscopy (fig.~\ref{fig:sf28}). The  fragmentation of states is not very strong and  the main part of the spectroscopic force is concentrated in the corresponding first excited states. In case of $^{30}$Si the picture is more complicated. Spectroscopy of $^{30}$Si from one-nucleon transfer reaction, adopted in the joint numerical analysis is shown in fig.~\ref{fig:sf30}. The state $1d_{5/2}$ in reaction of neutron pickup is fragmented significantly  without a pronounced maximum of spectroscopic strength. This circumstance leads to additional uncertainty in the evaluation of $1d_{5/2}^{\nu}$ single-particle energy.

 \begin{table}
\caption{Proton single-particle energies $-E_{nlj}$ (in MeV) in even Si isotopes}
\label{tab:2}       
\begin{center}
\begin{tabular}[t]{crrrr}
\hline\noalign{\smallskip}
A & $1d_{5/2}$ & $2s_{1/2}$ & $1d_{3/2}$ & $1f_{7/2}$ 
\\
\noalign{\smallskip}\hline\noalign{\smallskip}
${26}^{b)}$ & $ 5.51(3)$ & $ 2.76(3)$ & $- 0.25(3)$ & $- 2.58(3)$  \\

${28}^{a)}$& $ 9.68(2)$ & $ 6.42(1)$ & $ 3.70(2)$ & $- 0.70(1)$  \\

${30}^{a)}$ & $12.65(2)$ & $8.98(1)$ & $6.53(2)$ & $2.87(1)$  \\

${32}^{b)}$ &$ 15.54(2)$ &$ 11.11(2)$ & $9.40(2)$ & $5.32(1)$  \\

${34}^{b)}$ & $17.76(2)$ & $12.93(2)$ & $10.87(2)$ & $8.09(2)$  \\

${36}^{b)}$ & $18.80(8)$ & $13.89(8)$ & $13.03(8)$ &   \\

${38}^{b)}$ & $20.7(2)$ & $16.0(2)$ & $15.6(2)$ &   \\

${40}^{b)}$ & $22.6(6)$ & $17.7(6)$ & $17.5(6)$ &   \\

${42}^{b)}$ & $23.9(10)$ & $19.2(10)$ & $19.0(10)$ &   \\
\noalign{\smallskip}\hline
\multicolumn{5}{p{7.5cm}}{$^{a)}$ {\small Single-particle energies from joint analysis of one-nucleon transfer
reaction data. Statistical errors are presented. For systematic uncertainties see the text.}}\\
\multicolumn{5}{p{8cm}}{\small{$^{b)}$ Evaluations  are based on odd nuclei spectra. Errors in parentheses are from uncertainties in masses. }}\\
\end{tabular}
\end{center}
\end{table}

\begin{table}
\caption{Neutron single-particle energies $-E_{nlj}$ (in MeV) in even Si isotopes}
\label{tab:3}       
\begin{center}
\begin{tabular}[t]{crrrrr}
\hline\noalign{\smallskip}
$A$ &  $1d_{5/2}$ & $2s_{1/2}$ & $1d_{3/2}$ & $1f_{7/2}$ & $2p_{3/2}$\\
\noalign{\smallskip}\hline\noalign{\smallskip}
${26}^{b)}$ & $17.13(2)$ & $12.53(1)$ & $10.45(1)$ &  & \\

${28}^{a)}$ &  $15.45(2)$ & $12.07(1)$ & $9.62(2)$ & $3.47(3)$ & \\

${30}^{a)}$ &  $14.22(5)$ & $9.36(2)$ & $8.10(1)$ & $2.26(4)$ & \\

${32}^{b)}$ & $11.99(1)$ & $9.95(1)$ & $6.85(1)$ & $3.07(1)$ & $2.53(1)$ \\

${34}^{b)}$ &  $11.86(2)$ & $8.52(2)$ & $7.51(2)$ & $2.47(4)$ & $1.57(4)$ \\

${36}^{b)}$ &  $8.27(8)$ & $7.79(8)$ & $7.08(8)$ & $3.2(1)$ & $1.6(1)$ \\

${38}^{b)}$ &  & $7.2(1)$ & $6.4(1)$ & $3.6(2)$ &   \\

${40}^{b)}$  &  &  & $5.8(3)$ & $4.1(5)$ & \\

${42}^{b)}$  &  &  &  & $3.6 (10)$ & \\
\noalign{\smallskip}\hline
\multicolumn{6}{p{8cm}}{$^{a)}$ {\small Single-particle energies from joint analysis of one-nucleon transfer
reaction data. Statistical errors are presented. For systematic uncertainties see the text.}}\\
\multicolumn{6}{p{8cm}}{\small{$^{b)}$ Evaluations  are based on odd nuclei spectra. Errors in parentheses are from uncertainties in masses. }}\\
\end{tabular}
\end{center}
\end{table}

The proton and  neutron  single-particle energies (\ref{enlj}) for $^{28}$Si and $^{30}$Si are presented in Tabl.~\ref{tab:2} and Tabl.~\ref{tab:3}, respectively. 
The corresponding occupation probabilities $N_{nlj}$ for protons and neutrons (\ref{nnlj}) were estimated as well. Results are presented  in Tabl.~\ref{tab:4}. In the  isotope $^{28}$Si the total occupation of the proton $2s1d$ shell ${N_{1d2s}=5.8}$ and the neutron one ${N_{1d2s}=6.2}$ in agreement with the IPM predictions. The values of the Fermi energy  $E_F^{\pi}$ = $-7.6 \pm 0.1$~MeV for protons and 
${E_F^{\nu}  = -12.9 \pm 0.8}$~MeV for neutrons  as well as values of ${\Delta^{\pi} = 4.0 \pm 0.2}$~MeV and \\ ${\Delta^{\nu} = 3.5 \pm 1.5}$~MeV (\ref{BCS}) agree with the corresponding values based on the atomic masses.

Otherwise, in case of the $^{30}$Si isotope occupation of the proton $1d2s$ shell is overestimated and its value is $N_{1d2s}=6.09 \pm 0.16 $. Addition of two neutrons causes gradual occupancy of the $1f_{7/2}$ shell. Taking $1f_{7/2}$ neutrons into account ($N_{nlj}(1f_{7/2}) = 0.06$) would lead to the total $N_{1d2s}=7.8\pm 0.2$ for neutrons. Its correspondence to $N_{1d2s}=8$ predicted by the IPM  is accompanied by insignificant difference (0.5~MeV) between $E_F^{\nu}  = -9.1 \pm 0.3$~MeV  and  Fermi energy (\ref{EF}) from experimental masses. The same difference for proton Fermi energy $E_F^{\pi} = -10,07 \pm 0.02$~MeV is comparable to that in $^{28}$Si. Gap parameters $\Delta^{\pi} = 2.34\pm 0.03$~MeV and $\Delta^{\nu} = 1.8 \pm 0.8$~MeV are also in good agreement with values based on the atomic masses. 

\begin{table}
\caption{Proton and neutron occupation probabilities $N_{nlj}$ of $^{28,30}$Si isotopes}
\label{tab:4}       
\begin{center}
\begin{tabular}{ccccc}
\hline\noalign{\smallskip}
&\multicolumn{2}{c}{\small{$^{28}$Si}} & \multicolumn{2}{c}{\small{$^{30}$Si}}\\
\noalign{\smallskip}\hline\noalign{\smallskip}
Subshell
& $\pi$ & $\nu$ & $\pi$ & $\nu$ \\
\noalign{\smallskip}\hline\noalign{\smallskip}
$1d_{5/2}$ & 0.75(07) & 0.77(07) & 0.87(07) & 0.85(06)\\

 $2s_{1/2}$ & 0.37(12) & 0.40(10) & 0.29(11) & 0.58(12)\\

 $1d_{3/2}$ & 0.15(10) & 0.21(10) & 0.08(09) & 0.25(09)\\

 $1f_{7/2}$ &  &  &  & 0.06(07)\\
\hline\noalign{\smallskip}
\end{tabular}
\end{center}
\end{table}

Statistical errors for $^{28}$Si and $^{30}$Si data presented in parentheses are from one-nucleon transfer reaction. The errors were estimated in proposition of 20\% uncertainty of large value spectroscopic factors $(C^2S^{\pm}_{nlj}>1)$ and 50\% uncertainty for $(C^2S^{\pm}_{nlj}<1)$ \cite{CP74}. Nevertheless, the statistical errors are much smaller than systematic uncertainties tied with the method of  joint numerical analysis of stripping and pickup experiments data. If the range of considered experiment pairs would be limited by the value of penalty function less than 0.3, the single-particle energy deviations achieve  $+2 \div -1.5$~MeV. At the same time the value of Fermi energy $E_F$ does not differ from the experimental value (\ref{EF}) by more than 1~MeV. It can be suggested that criteria used allow to determine the  $1d2s$ shell structure with accuracy no more than 10 -- 15\%.

In Tabl.~\ref{tab:4} the occupation probabilities for $1d2s$ shell in $^{30}$Si are presented. Addition of two neutrons leads to the significant increase of occupancy of the proton subshell $1d_{5/2}$. On the other hand, the occupancy of $2s_{1/2}$ decreases causing the slight depletion of the central proton density.
However, for the even-even exotic isotopes $^{26}$Si and $^{32-42}$Si spectroscopic data are not available, but the proton $N_{nlj}$ values obtained for the stable isotopes may be used in further evaluation of the single-particle energies for the neutron excess isotopes.

\subsection{Proton single-particle states in unstable Si isotopes }
\label{sec:2:2}

The pickup and stripping spectroscopies for the stable isotopes $^{28}$Si and $^{30}$Si prove that the most of spectroscopic strength is concentrated in the first excited state with the corresponding spin-parity $J^{\pi}$ with the only exception -- the highly fragmented ${5/2}^+$ state. Thus, evaluation of the single-particle energies based on the spectra of  adjoining odd nuclei \cite{Bohr} is reasonable as an initial rough estimation. 

The energy disposition of the subshell depends significantly on the ratio of particles and holes in it, i.e. on the corresponding occupancy. In the framework  of the closed $1d_{5/2}$ and vacant $2s_{1/2}$ and $1d_{3/2}$ subshells the result essentially disagrees with the joint analysis data. In the case of $^{28}$Si the approximate maximum value of this disagreement is 3~MeV for the $1d_{3/2}$ subshell. Taking the experimental occupation probabilities (Tabl.~\ref{tab:4}) into account would improve the results significantly by minimising the disagreement (0.8~MeV). The same situation occurs for $^{30}$Si; although the difference between the rough estimation and the joint analysis data is not that remarkable, application of the occupation probabilities would improve the correlation of two approaches as well.

As it was shown in sec.~\ref{sec:2:1}, two additional neutrons in $^{30}$Si relatively to $^{28}$Si cause the noticeable increase of $1d_{5/2}$ occupancy and its decrease in $2s_{1/2}$. Assuming this tendency remains throughout the $^{32-42}$Si isotopes, one may apply the experimental occupation probabilities in $^{30}$Si as the same values in the exotic isotopes in order to evaluate the proton single-particle energies. The results are presented in Tabl.~\ref{tab:2}. The occupation probability of $1d_{5/2}$ subshell in all neutron-rich isotopes $^{32-42}$Si  was assumed to be $N_{nlj}(1d_{5/2})=0.9$. Model calculations lead to a somewhat smaller value, however, as the number of neutrons increases, the $1d_{5/2}$ occupancy grows and reaches the value 0.87 in isotope $^{40}$Si (see fig.~\ref{fig:po} for model results). It is almost impossible to estimate the values of $N_{nlj}(2s_{1/2})$ and $N_{nlj}(1d_{3/2})$ on the basis of experimental data. To maintain a smooth tendency to decrease, the  occupation probabilities of $2s_{1/2}$ and $1d_{3/2}$ subshell in $^{32-34}$Si were defined as $N_{nlj}(2s_{1/2})=0.2$ and $N_{nlj}(1d_{3/2})=0.1$. For the rest of the isotopes $^{36-42}$Si the $E_{nlj}$ values for $2s_{1/2}$ and $1d_{3/2}$ states was based entirely on the neighbouring odd $A+1$ nuclei to represent the stripping reaction mechanism.

\subsection{Neutron single-particle states in unstable Si isotopes }
\label{sec:2:3}
 In the scope of this approach one must pay particular attention to the centroid evaluation for the neutron levels in the selected Si isotopes. The adjacent odd isotopes' spectra are highly informative for this purpose as well. As it was proved for the isotopes with  magic $Z$ number (e.g. Ca and Sn isotopes) the single states reveal particular behaviour: the excitation energy of a level corresponding to a state above the Fermi energy decreases up to zero value with the neutron number increasing. As long as this shell is filled, this state  remains the ground state, then the excitation energy starts to grow smoothly.

A similar situation can be observed in odd Si isotopes. However, probable splitting caused by the deformed potential, the states of rotational and vibrational origin make it more uncertain and vague. Information on the excited states is available only for $^{25-39}$Si, as long as the spin-parity of the only excited state in $^{41}$Si has not been identified yet \cite{ENSDF}.
  
 The excited-state spectrum for $^{25}$Si is too scanty to define the single-particle states in $^{26}$Si. The most sufficient and reliable information on the structure of $^{25}$Si was gathered in the reaction $^9$Be($^{26}$Si,$^{25}$Si)X \cite{Rey}. Comparison of the experimental results and the shell model predictions proved the first excited state $E_1(1/2^+) = 0.821$~MeV to be one of the rotational levels. However, due to the lack of spectrum data, only this state can be used to estimate $C_{nlj}^{-}$ and $E_{nlj}^{-}$ for $2_{1/2}$ in $^{25}$Si. In order to define the stripping values $C_{nlj}^{+}$ and $E_{nlj}^{+}$ the spectrum for $^{27}$Si was used. There are many excited states with $J^{\pi} = {3/2}^+$ and the first one has  energy $E(3/2_1^+)$ too low to be of the single-particle origin. The energy of the second state $E(3/2_2^+) = 2.866$~MeV  in $^{27}$Si can be traced by the corresponding analogue state in $^{27}$Al  and this state and along with the first excited states in $^{29-35}$Si reveal the typical single-particle behaviour mentioned above. For draft estimations  the $^{26}$Si isotope occupation probabilities were chosen according to the IPM values $N_{nlj}(d_{5/2})=2/3$ and $N_{nlj}(2s_{1/2}) = N_{nlj}(1d_{3/2})=0$. The corresponding single-particle energies in $^{26}$Si are presented in Tabl.~\ref{tab:3}.
 
The stable deformation of Si isotopes makes the evaluation based on spectra of odd nuclei more rough than the results of joint analysis of one-nucleon transfer reaction data. On the other hand, in order to estimate the discrepancy between these two approaches, this evaluation was made for stable isotopes $^{28, 30}$Si. In this case of $^{28}$Si the spectrum of $^{27}$Si was used to define pickup values of $C_{nlj}^{-}$ and $E_{nlj}^{-}$. 

The energy dispositions of  $1d_{5/2}$ and $2s_{1/2}$ states are based on the $E_{g.s.}$ and $E(1/2_1^+) = 0.781$~MeV. As it was already mentioned above, $E_2(3/2^+)$ was used in case of the $1d_{3/2}$ state. Since $^{28}$Si reveals low occupation of $1f_{7/2}$ state, pickup spectrum of $^{27}$Si cannot be used to define $E_{nlj}(1f_{7/2})$, the only $E(7/2_1^-) = 7.7$~MeV is essentially higher than other states and can be barely occupied.

The spectrum of $^{29}$Si would correspond to both the resulting stripping spectrum for the $^{28}$Si target and pickup spectrum for $^{30}$Si. The excited states in this spectrum are distinctly isolated up to 4.6~MeV, thus the first excited states were used for the evaluation. The same was done to evaluate $C_{nlj}^{+}$ and $E_{nlj}^{+}$ for $^{30}$Si on the base of $^{31}$Si spectrum, except for the $1d_{5/2}$ state which was estimated by using the excitation energy of the second state $E(5/2_2^+) = 2.788$~MeV.

Neutron single-particle states of $2s1d$ subshell for the  isotope $^{28}$Si evaluated with the IPM-based occupation probability $N_{nlj}(1d_{5/2}) = 1$, $N_{nlj}(2s_{1/2}) = N_{nlj}(1d_{3/2}) = 0 $ differ significantly from the experimental values with the maximum difference $2.5 - 2.9$~MeV for $2s_{1/2}$ and $1d_{3/2}$ subshells. Taking the experimental values of the occupation probabilities from Tabl.~\ref{tab:4} into account decreases this difference to 1.5~MeV. The same result is obtained for  isotope $^{30}$Si. Strong difference for $2s_{1/2}$ and $1d_{3/2}$ states can be eliminated by utilizing the $N_{nlj}$ values. In doing so, however, the discrepancy for $1d_{5/2}$ state rises up to  $\sim 2.9$~MeV. These facts point to definite misinterpretation of the single-particle spectra: the states fragmentation (especially for $5/2^+$ state) must be taken into account. It should be noted that, despite the fact that the errors in nuclear mass determination are very small, the uncertainty of this method is very significant. Comparison of the evaluation based on the spectra of odd nuclei with the results of joint analysis gives rise to put an 20\% uncertainty in single-particle energy estimations.   

For the $^{32}$Si isotope the spectrum of $^{33}$Si was used to define the stripping values of $C_{nlj}^{+}$ and $E_{nlj}^{+}$. The majority of $J^{\pi}$ spin-parities are presented by the single states. The values
${N_{nlj}(1d_{5/2}) = N_{nlj}(2s_{1/2}) =1}$,  $N_{nlj}(1d_{3/2}) = 1/2$ were used for the evaluation of the single-particle energies. Result is demonstrated in Tabl.~\ref{tab:3}.

 The neutron structure of isotopes with ${N \ge 20}$  is based on the spectra of $^{35-41}$Si. The occupation of  neutron $2s1d$ shell causes dramatic changes in these spectra. First of all, density of the available states up to 2~MeV increases. Secondly, the energy position of each state does not vary essentially with the increasing $N$. Thirdly, the energy position of $5/2^+$ state decreases while moving from $^{35}$Si to $^{37}$Si. The states of positive spin-parity may be assumed to be of  collective origin and related with the $(3/2^+)_{\nu}^-$ configuration. Thus, the available data do not allow to obtain any appropriate energy value for  $1d_{5/2}$ state as long as the $2s1d$ shell is closed. In the framework of the closed $2s1d$ shell and gradually occupied $1f_{7/2}$ state the single-particle energies were evaluated as well (see Tabl.~\ref{tab:3}). According to the main tendency in the spectra, the neutron single-particle energies are stabilized and  get closer approaching the  isotope $^{42}$Si.

\section{Description of the model and parametrizations of the potential}
\label{sec:3}
 The DOM mean field is unified both for negative and positive energies, so it permits one to calculate single-particle characteristics of nucleus as well as nucleon scattering cross section.

The nucleon-nucleus optical model local equivalent potential is given as:
$$U(r,E)=V_{C}(r)-U_{p}(r,E)-U_{SO}(r,E),$$
where $V_{C}(r)$ 
is the Coulomb potential (for protons), which is usually taken to be that of a uniformly charged sphere of radius $R_{C}=r_{C}A^{-1/3}$; $U_{p}(r,E)$ is the central part of the nucleus potential; $U_{SO}(r,E)$ is the spin-orbit potential, with the standard Thomas formfactor.

According to \cite{MS91} the central real part of the dispersive optical potential (DOP) is represented by the sum of three terms, namely, the Hartree-Fock (HF) type potential $V_{HF}$, the volume, $\Delta V_{s}$, and surface, $\Delta V_{d}$ dispersive components so that $U_{p}(r,E)$ is expressed by:
\begin{eqnarray*}
U_{p}(r,E)=V_{HF}(E)f(r,r_{HF},a_{HF})+\\
+\Delta V_{s}(E)f(r,r_{s},a_{s})-4a_{d}\Delta V_{d}(E)\frac{d}{dr}f(r,r_{d},a_{d})+\\
+iW_{s}(E)f(r,r_{s},a_{s})-4ia_{d}W_{d}(E)\frac{d}{dr}f(r,r_{d},a_{d}),     
\end{eqnarray*}
where $f(r,r_i, a_i)$ is the Saxon-Woods function.

The dispersive components are determined from the dispersion relation:
\begin{eqnarray*}
\Delta V_{s(d)}(E)=(E_{F}-E)\frac{P}{\pi}\int\limits_{-\infty}^{\infty}\frac{W_{s(d)}(E')}{(E'-E_F)(E-E')}dE' ,
\end{eqnarray*}
 where $P$ denotes the principal value of the integral. The components $\Delta V_{s,d}(E)$ vary rapidly in the Fermi energy region
and lead to the grouping of single-particle energies  of valence states close to the energy $E_F$. These components represent the individual features of the nucleus.

The energy dependence of the imaginary potential depth at $E, E_0>E_F$ was defined in the present paper as:
$$W_{d}(E)=d_{1}\frac{(E-E_{0})^2 \exp[-d_2(E-E_0)]}{(E-E_0)^2 +d_3^2} ,$$
for the surface component and 
$$W_{s}(E)=w_1\frac{(E-E_{0})^2}{(E-E_0)^2 +w_2^2} ,$$
for the volume component.
The imaginary part of DOP was assumed to be symmetric relatively to the Fermi energy and equal to zero in the range $|E-E_F|<|E_0-E_F|$. We have neglected the symmetry breaking that arises because of the non-locality of the imaginary part of the DOP, since it has little effect on single-particle energy levels.

\begin{table}
\caption{The parameters of DOPs (in MeV) of the Si isotopes. See text for the details. }
\label{tab:3p}       
\begin{center}
\begin{tabular}[t]{crrrrrr}
\hline\noalign{\smallskip}
Isotope & \multicolumn{2}{c}{$-E_F$} & \multicolumn{2}{c}{$-E_0$} & \multicolumn{2}{c}{$V_{ HF}(E_F)$} \\
\noalign{\smallskip}\hline\noalign{\smallskip}
&$\pi$ & $\nu$ &$\pi$ & $\nu$ &$\pi$ & $\nu$ \\
\noalign{\smallskip}\hline\noalign{\smallskip}
$^{26}$Si & 3.2  & 16.2 & 3.2	& 16.2	& 52.5 &  61.4\\
$^{28}$Si & 7.2 & 12.8 & 6.0 &  11.0& 56.8 	& 56.5\\
$^{30}$Si & 10.4 & 8.6 &10.4	&  6.0 & 59.8  & 51.0\\
$^{32}$Si & 13.0 & 6.9 & 13.0 & 6.9	& 61.6 &   49.6\\
$^{34}$Si & 15.5 &5.0 & 11.0	&  2.0 & 63.5  & 48.0\\
$^{36}$Si & 16.7 & 4.2 &16.7	&  4.2  & 63.4 	& 48.2\\
$^{38}$Si & 18.6 & 3.6 &18,6	&  3.6  & 64.4 	& 48.4\\
$^{40}$Si & 20.3 & 3.1 & 20.3 &  3.1 & 65.5 	& 47.3\\
$^{42}$Si & 21.9 & 2.6 & 21.9	&  2.6 & 66.1 	& 46.0\\

\noalign{\smallskip}\hline

\end{tabular}
\end{center}
\end{table}

The energy dependence of the HF potential depth was parametrized by the exponential function:
$$V_{HF}(E)=V_{HF}(E_F)\exp\left({-\frac{\gamma (E-E_F)}{V_{HF}(E_F)}}\right) .$$

Single-particle energies $E_{nlj}$ and the total wave functions $\Phi_{nlj}(\vec{r})$ were calculated by solving the Schrodinger equation with the real part of the DOP in iteration procedure.
Further the density distributions were calculated by the single-particle approach:
\begin{equation}
\rho_{p(n)}(r)=\frac{1}{4\pi}\sum_{nlj}(2j+1)N_{nlj}\bar{u}_{nlj}^2(r).
\label{ropn}
\end{equation}
The radial part ${u}_{nlj}(r)$ of the total wave function $\Phi_{nlj}(\vec{r})$ was corrected to take the effect of non-locality into account:
\begin{equation}
\bar{u}_{nlj}(r)=\left(\frac{m^*_{HF}(r,E)}{m}\right)^{1/2}u_{nlj}(r), 
\label{ur}
\end{equation}
and then was normalized to unity. 
The ratio of the Hartree-Fock effective mass $m^*_{HF}(r,E)$ to the mass $m$ of the free nucleon is defined as:
$$\frac{m^*_{HF}(r,E)}{m}=1-\frac{d}{dE}V_{HF}(r,E).$$
Following \cite{JMN85}, transition from the proton density to the charge density was performed by using the relation:
$$\rho_{ch}(r)=(\pi a^2)^{-3/2}\int \rho_p(r')e^{-(r-r')^2/a^2}dr', $$
where  the value $a^2$ = 0.4~fm$^2$ approximately takes  the proton charge form factor and
the centre-of-mass motion into account.

 \begin{figure}
\resizebox{9 cm}{!}{%
\includegraphics{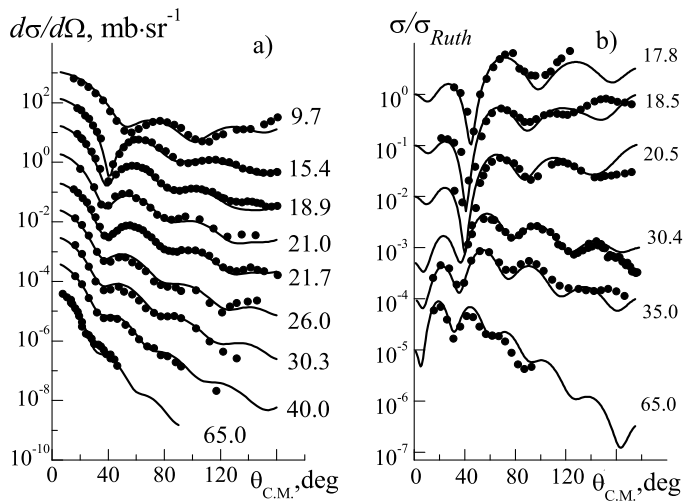}
}
\caption{Neutron (a) and proton (b) elastic differential cross sections for $^{28}$Si. Experimental data \cite{AO12,Ex1,Ex2,Ex3,Ex4,Ex5,Ex6,Ex7,Ex8,Ex9,Ex10} are shown by circles, the predictions are DOM calculation (lines). }
\label{fig:1}       
\end{figure}

Occupation probabilities $N_{nlj}$ in (\ref{ropn}) were calculated in two different ways. According to the first way,  probabilities  $N_{nlj}$ were determined by using expression (\ref{BCS}) of BCS theory
with the empirical pairing gap parameter $\Delta$ (\ref{Dn}) calculated from nuclear masses data \cite{AME12}, and the energies $E_{nlj}$ obtained by DOM. 
In the second way, probabilities $N_{nlj}$ were determined by approximate formulae of the DOM \cite{MS91}: \\
\begin{equation}
\begin{array}{ll}
N_{nlj}=&1-\int\limits_0^{\infty}\bar{u}_{nlj}^2(r)\Bigl[\{m^*_{HF}/m(r,E_{nlj})\}^{-1}\times \\
&\times{\pi}^{-1}\int\limits_{E_F}^{\infty}\frac{W(r,E')}{(E'-E_{nlj})^2}dE'\Bigr]dr, 
 \mbox{ } E_{nlj}<E_F, \\ 

N_{nlj}=&\int\limits_0^{\infty}\bar{u}_{nlj}^2(r)\Bigl[\{m^*_{HF}/m(r,E_{nlj})\}^{-1}\times\\  
&\times{\pi}^{-1}\int\limits_{-\infty}^{E_F}\frac{W(r,E')}{(E'-E_{nlj})^2}dE'\Bigr]dr,\mbox{ } E_{nlj}>E_F.
\end{array}
\label{NDOM}
\end{equation}
For deep lying states, occupation probabilities $N_{nlj}$ (\ref{BCS}) approach 1 in contrast to $N_{nlj}$ (\ref{NDOM}) which are approximately 10\% less due to the short-range correlations. The latter push single-particle strength from orbits of IPM out to the energies of hundreds of MeV \cite{DB04}. Occupation probabilities $N_{nlj}$ (\ref{NDOM}) agree with the results of $(e, e'p)$ measurements \cite{Lap93}. At the same time, the agreement of total number of neutrons (protons) in bound states $N_{n(p)} = \Sigma (2j + 1) N_{nlj}$ with $N(Z)$ of nucleus is achieved within 10\% if the occupation probabilities $N_{nlj}$ (\ref{NDOM}) are used. For example, occupation probabilities $N_{nlj}$ of the proton bound states in $^{90}$Zr with Z = 40 \cite{Wang93} correspond to the total number of protons $N_p$ = 37.3. Values of $N_{nlj}$ from \cite{AO12} lead to the total numbers $N_n$ = 13.8, $N_p$ = 12.8 in the bound states of $^{28}$Si. The correct number of particles was achieved in dispersive self-energy method \cite{Dick17} and references therein. It was shown that the resulting substantial excess of the number of particles was the consequence of incorrect normalization when local-equivalent energy-dependent real potential was used in  Dyson equation. The proper number of particles was obtained due to returning to a fully non-local form of the potential. The local-equivalent energy-dependent real part of the potential distorts also the unitarity of the single-quasiparticle dispersive optical model \cite{Kol14}. The method to restore the unitarity of the model was proposed in \cite{Kol17}. The method lead to the conservation of the particle number and, in neglect of pairing, to the equality of $N_{nlj}$ to 1 and 0 for the states with $E_{nlj} < E_F$ and $E_{nlj} >  E_F$ respectively.  The correct number of particles $N_{n(p)}$  in bound states is achieved also if occupation probabilities $N_{nlj}$ (\ref{BCS}) and energies $E_{nlj}$ calculated by DOM \cite{MS91} are used.

Spectroscopic factors $S_{nlj}$ relative to Independent Particle Model, were calculated by the formula:
\begin{equation}
S_{nlj}=\int\limits_0^{\infty}\bar{u}_{nlj}^2(r)\Bigl[\frac{m}{\bar{m}(r,E_{nlj})}\Bigr]dr,
\end{equation}\label{Snlj}
where $\bar{m}(r,E)$ is the energy-dependent effective mass, the so called {"E-mass"}.  The mass  is connected with the dispersive component of DOP by the expression:
 \begin{equation}
\frac{\bar{m}(r,E_{nlj})}{m} = 1 - \Bigl[\frac{m}{m^*_{HF}(r, E)} \Bigr]\frac{d}{dE}\Delta V(r,E) \end{equation}\label{M_e}

\section{Calculation results}
\subsection{DOP parameters and the nucleon scattering cross sections for  $^{28}$Si}
\label{sec:3a}

\begin{figure}
\resizebox{9 cm}{!}{%
\includegraphics{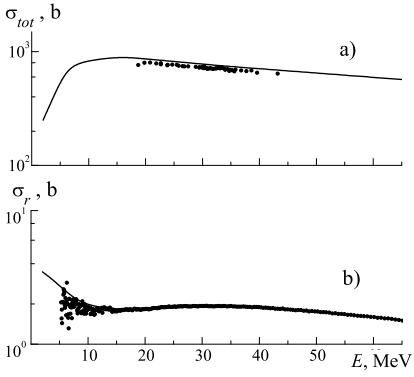}
}
\caption{Neutron total interaction (a) and proton total reaction (b) cross sections for $^{28}$Si. Circles - experimental data \cite{FAF93,Ex11}, lines - DOM calculation.}
\label{fig:2}      
\end{figure}

We determined the DOP parameters for the stable isotope
 $^{28}$Si, for which the experimental data on proton and neutron scattering cross sections exist. The volume $W_s$ and surface $W_d$ imaginary parts of DOP as well as spin-orbit and Coulomb potentials were fixed according to the global parameters \cite{KD03}. Geometrical parameter $r_{HF}$ was chosen to be equal to $r_v=1.23$~fm \cite{AO12}. In this paper, neutron scattering data and single-particle energies for $^{28}$Si were analyzed by DOM. It was found that the experimental total interaction cross sections $\sigma_{tot}$ can be described with the radius parameter $r_{HF}$ of Hartree-Fock component, which increases from 1.18 to 1.23~fm as energy increases from 0 to 25~MeV. We were also faced with the fact that neutron cross sections $\sigma_{tot}$ which were calculated with $r_{HF} = r_v = 1.17$ fm \cite{KD03} underestimates the experimental data \cite{FAF93} at $E > 15$~MeV but leads to good agreement at $E <15$~MeV. As we assumed,  deformation can contribute to the deviation of the spherical DOM results from the experimental data. Notice that the energy dependence of the DOM geometrical parameters was also obtained in \cite{WF92,R00}. Thus, for the system $d+^{208}$Pb, parameters $a_{HF}$, $r_d$, and $a_d$ demonstrate the pronounced energy dependence \cite{WF92}. For the system $p+^{90}$Zr, the energy dependence was also introduced in the parameters $r_d$, and $a_d$ to describe low energy experimental data on total reaction cross sections $\sigma_r$ \cite{R00}.

After the fixing of $r_{HF} = 1.23$~fm for both neutrons and protons, parameter $V_{HF}(E_F) = 56.6$, $56.9$~MeV was determined  for neutrons and protons respectively from the fitting to Fermi energy. Parameter $\gamma= 0.420$ was chosen from the agreement with the experimental data on $E_{nlj}$ \cite{WP84} within their uncertainty and it was fixed in further calculations  for another Si isotopes.

Calculated neutron and proton elastic scattering cross sections, neutron total interaction cross sections, and proton total reaction cross sections for $^{28}$Si are shown in Fig.~\ref{fig:1} and Fig.~\ref{fig:2}.  Agreement with experiment data is  close to that achieved by using the global parameters \cite{KD03}.

Similarly to the $^{28}$Si isotope, parameters of the imaginary part of DOP, spin-orbit and Coulomb potential were equaled to the global parameters \cite{KD03}, $r_{HF} = r_s  = 1.23$~fm, $a_{HF} = a_V$ \cite{KD03}, for all of even $^{26-42}$Si isotopes. The parameter $V_{HF}(E_F)$ as well as energies $E_F$ and $E_0$ of the neutron and proton DOPs are presented in Tabl.~\ref{tab:3p}. 

\begin{figure}
\resizebox{8 cm}{!}{%
\includegraphics{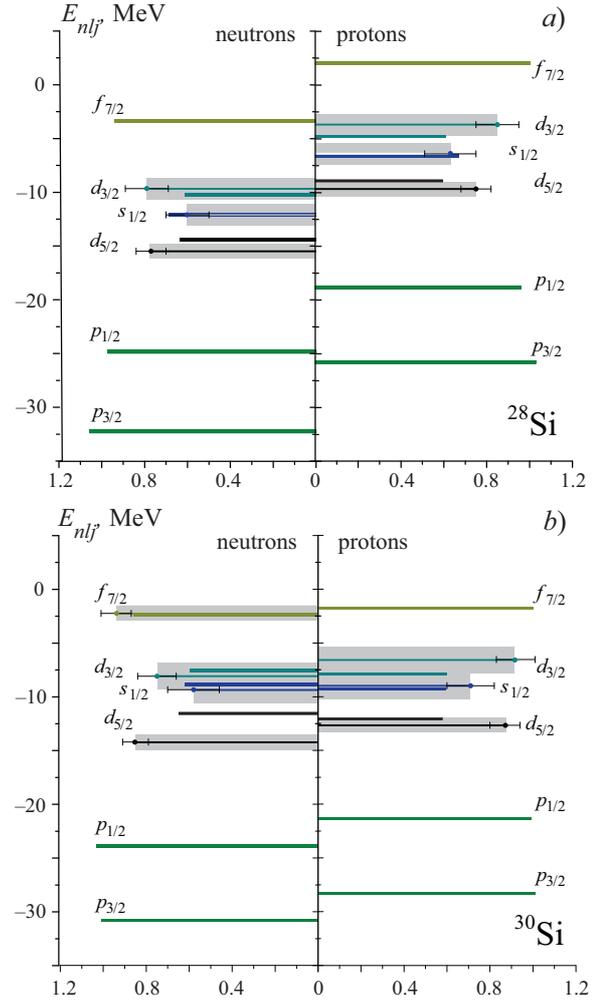}
}
\caption{(Colour online) Spectroscopic factors of quasiparticle states in $^{28}$Si (a) and $^{30}$Si (b). DOM results are displayed by filled bars, line bars with shadowed areas correspond to evaluated data. The error bars and shadowed region reflect the evaluation uncertainty (see the text). }
\label{fig:SF-DOM}       
\end{figure}

\subsection {Evolution of single-particle characteristics of even $^{26-42}$Si isotopes}
\label{sec:3b}

Spectroscopic factors $S_{nlj}$ calculated by DOM are shown in Fig.~\ref{fig:SF-DOM} in comparison with the evaluated data for $^{28,30}$Si isotopes. Good agreement within the error close to statistical uncertainty of evaluated data is observed for $2s_{1/2}$ and $1f_{7/2}$ states. For $1d$ states difference between calculated and evaluated spectroscopic factors $S_{nlj}$ is close to the total unsertainty of evaluated $S_{nlj}$. The lagest discrepancy is observed for neutron $1d_{5/2}$ state. Obtained agreement characterizes a degree of reliability of the evaluation for these isotopes.

The neutron and proton energies $E_{nlj}$ and occupation probabilities $N_{nlj}$ were calculated for even $^{26-42}$Si isotopes. The values of $N_{nlj}$ (\ref{NDOM}) lead to the total numbers $N_n$ = 13.9, $N_p$= 13.1 for $^{28}$Si. Neutron Fermi energy of $^{42}$Si isotope with N = 28 is close to 0, so that the states of the system $A+1$ are located at $E > 0$. For this isotope, total neutron number $N_n$ in bound states was found to be 25.7.  In this case the agreement of  $N_n$ with $N = 28$ improves if states with $E_{nlj} > 0$ are included in the sum. Sufficient agreement of calculated $E_{nlj}$ and $N_{nlj}$ (\ref{NDOM}) with the evaluated single-particle characteristics from Tabl.~\ref{tab:3} and Tabl.~\ref{tab:2} has been achieved for neutrons (Fig.~\ref{fig:no}) and protons ({Fig.~\ref{fig:po}). 
 
\begin{figure}
\resizebox{8.5 cm}{!}{%
\includegraphics{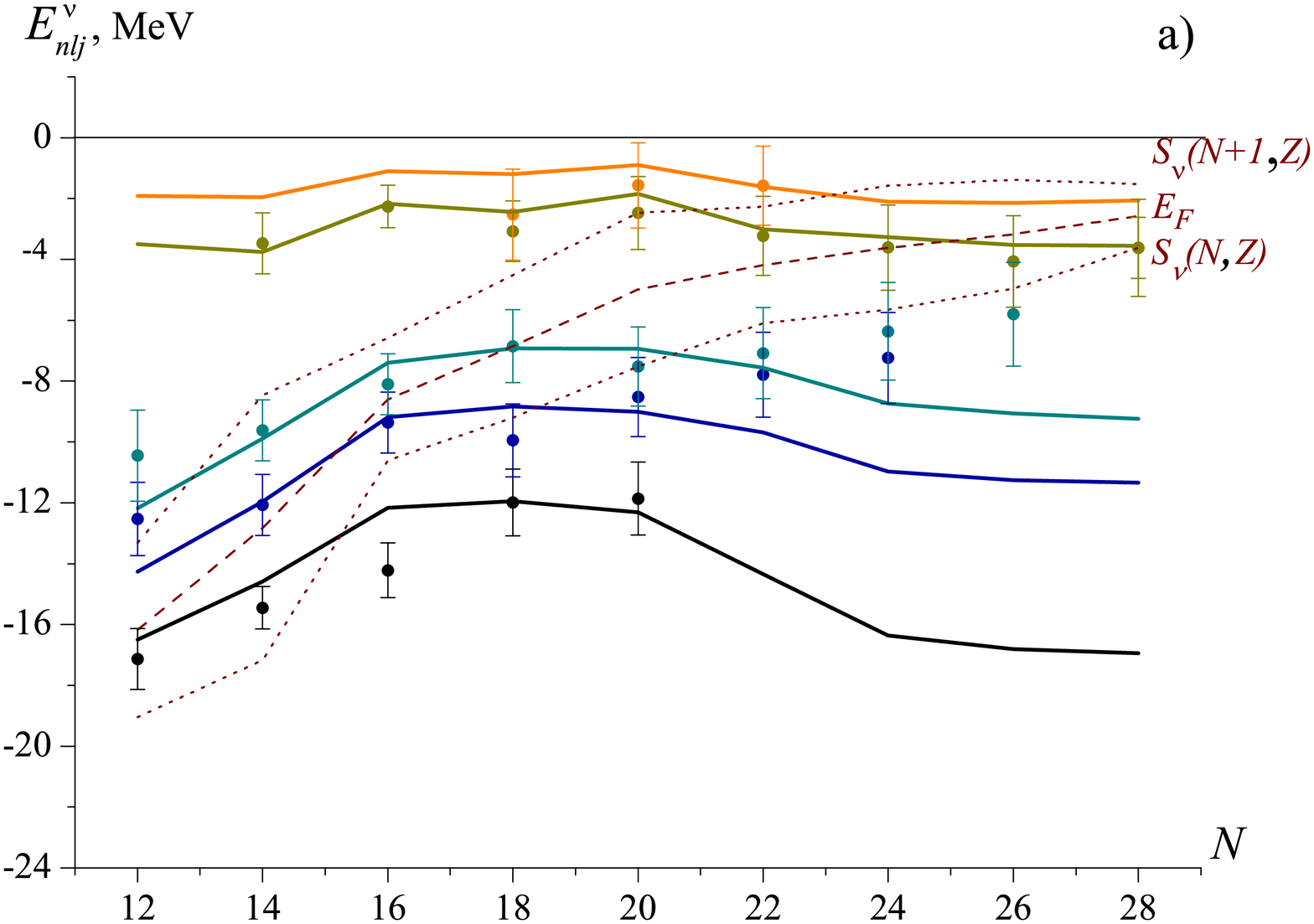}
}
\resizebox{8.5 cm}{!}{%
\includegraphics{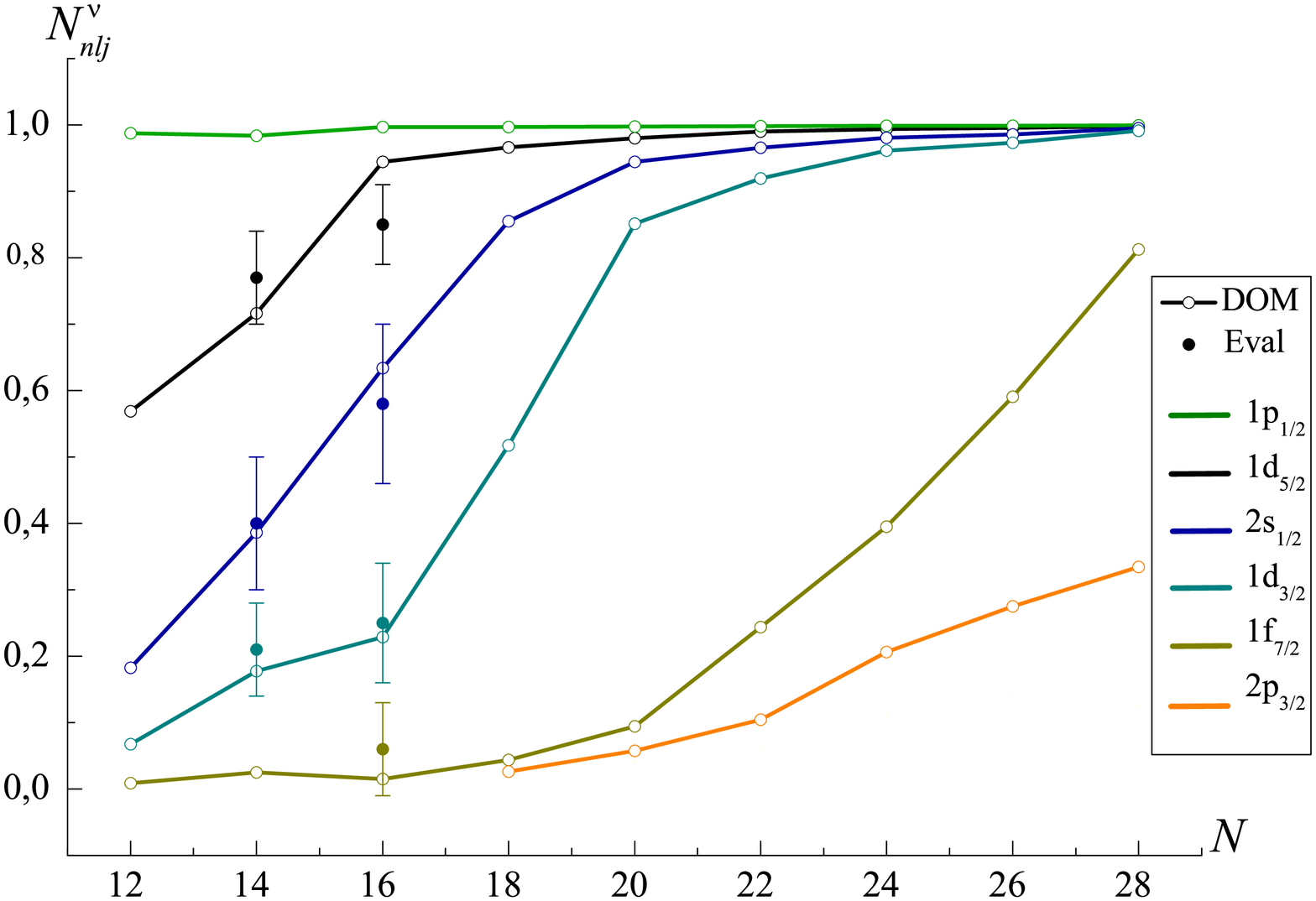}
}
\caption{(Colour online) Neutron single-particle energies $E^{\nu}_{nlj}$ (a) and occupation probabilities $N_{nlj}^{\nu}$ (b) of Si isotopes. The error bars reflect the evaluation uncertainty. }
\label{fig:no}       
\end{figure}

\begin{figure}
\resizebox{9 cm}{!}{%
\includegraphics{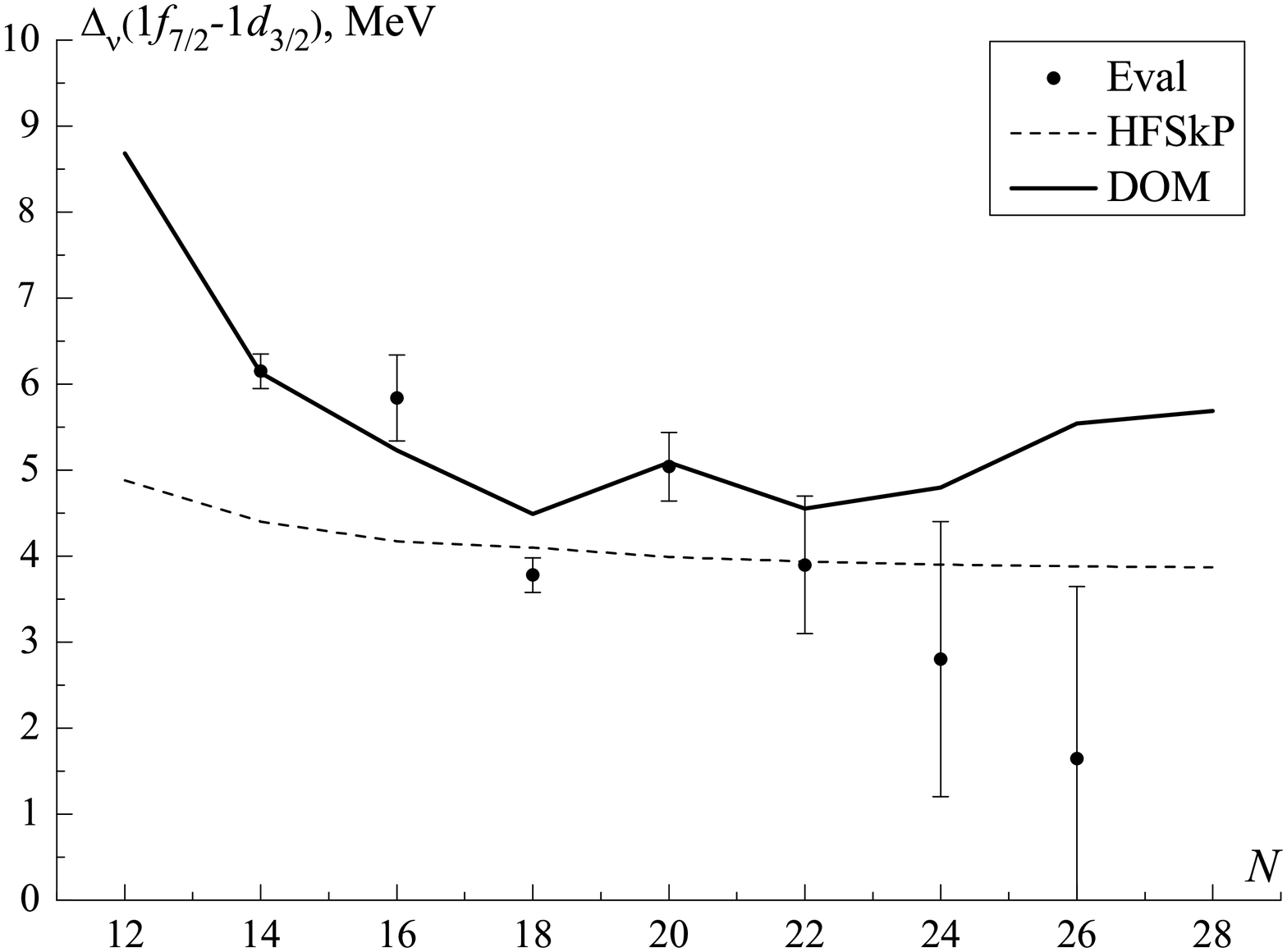}
}
\caption{Splitting between neutron single-particle energies of $1f_{7/2}$ and $1d_{3/2}$ states. }
\label{fig:fd_n}      
\end{figure}

In $^{28}$Si isotope with $N=Z$, neutron Fermi energy $E_F$ is located between the energies of $1d_{5/2}$ and $2s_{1/2}$ states, but these energies differ from the neutron separation energies (with the opposite sign) $ -S_n(^{28}\mbox{Si})$ and $ -S_n(^{29}\mbox{Si})$ \cite{WF92} respectively (see Fig.~\ref{fig:no}). In $^{30}$Si, energy $E_F$ is also located between the energies $E_{nlj}$ of $2s_{1/2}$ and $1d_{3/2}$ states, and these energies are close enough to separation energies $-S_n(^{30}$Si) and $-S_n(^{31}$Si) respectively. Such picture distinguishes $^{30}$Si among the neighbouring isotopes and corresponds with the increase of the excitation energy of the first $2^+$ state in $^{30}$Si isotope (see Fig.~\ref{fig:E2}). This reflects the magicity of $N = 16$ in Si isotopes. Magic properties of $N=20$ are completely revealed in neutron single-particle spectra and occupation probabilities of the Si isotopes. The energy gap between the last predominantly occupied $1d_{3/2}$ and the first predominantly unoccupied $1f_{7/2}$ states, evaluated as well as calculated by DOM, is wider in $^{34}$Si in comparison with neighbouring isotopes.  Corresponding calculated difference between $N_{nlj}$ for these states is equal to 0.77. Magic property of $N = 20$ number is also manifested in the widening of the gap between neutron $1f_{7/2}$  and $1d_{3/2}$ states (see Fig.~\ref{fig:no}) in $^{34}$Si. With further  increase of $N$, $1f_{7/2}$ state is occupied and its energy decreases. In $^{42}$Si with magic number $N=28$, the calculated particle-hole gap between  $1f_{7/2}$  and  $2p_{3/2}$ states is 1.5~MeV. This is essentially less than the experimental gap of $5.5\pm 1.1$~MeV in $^{48}$Ca \cite{BB05}. The difference between $N_{nlj}$ (\ref{BCS}) for $1f_{7/2}$ and  $2p_{3/2}$ states  equals 0.55 in $^{42}$Si. The reduction of particle-hole gap indicates the weakening of shell effects as the nucleus approaches the neutron drip line \cite{Do96} and agrees with the measurements of one-neutron knockout reaction from $^{36,38,40}$Si, which specify substantial neutron excitation across the $N = 28$ gap \cite{SG14}. The resulting 1.5~MeV gap is comparable with the difference of the neutron separation energies  ($S_n(N=28, Z=14) - S_n(N=29, Z=14)$)~=~2.1~MeV \cite{AME12}. It is substantially smaller than the estimation of about 4~MeV which is obtained from the experimental data \cite{Gau06}, assuming that the gap decreased evenly under the transition from $^{48}$Ca to $^{42}$Si. Experimental reduction amounts to about 330 keV per pair of protons removed from $^{49}$Ca. Calculation apparently indicates that the gap decreases unevenly.  The gap of 3.5~MeV in $^{42}$Si obtained by HF mean field model \cite{Peru00} can be treated as the upper limit since addition of the dispersive component to the HF component of the DOP  leads to the reduction of the particle-hole gap.

Fig.~\ref{fig:no} demonstrates the discrepancy between calculated and evaluated neutron energies of $1d_{3/2}$ and $2s_{1/2}$ states. The calculated energies slightly deepen with the increase of $N$ for $N > 20$, while the evaluated states become less bound, following Fermi energy. The discrepancy was probably caused by absence of experimental knowledge about fragmentation of these states. As a consequence, evaluated neutron $fd$ splitting between $1f_{7/2}$ and $1d_{3/2}$ states decreases, while calculated splitting increases  with $N$ varying from 20 to~28 (see Fig.~\ref{fig:fd_n}).

\begin{figure}
\resizebox{8.5 cm}{!}{%
\includegraphics{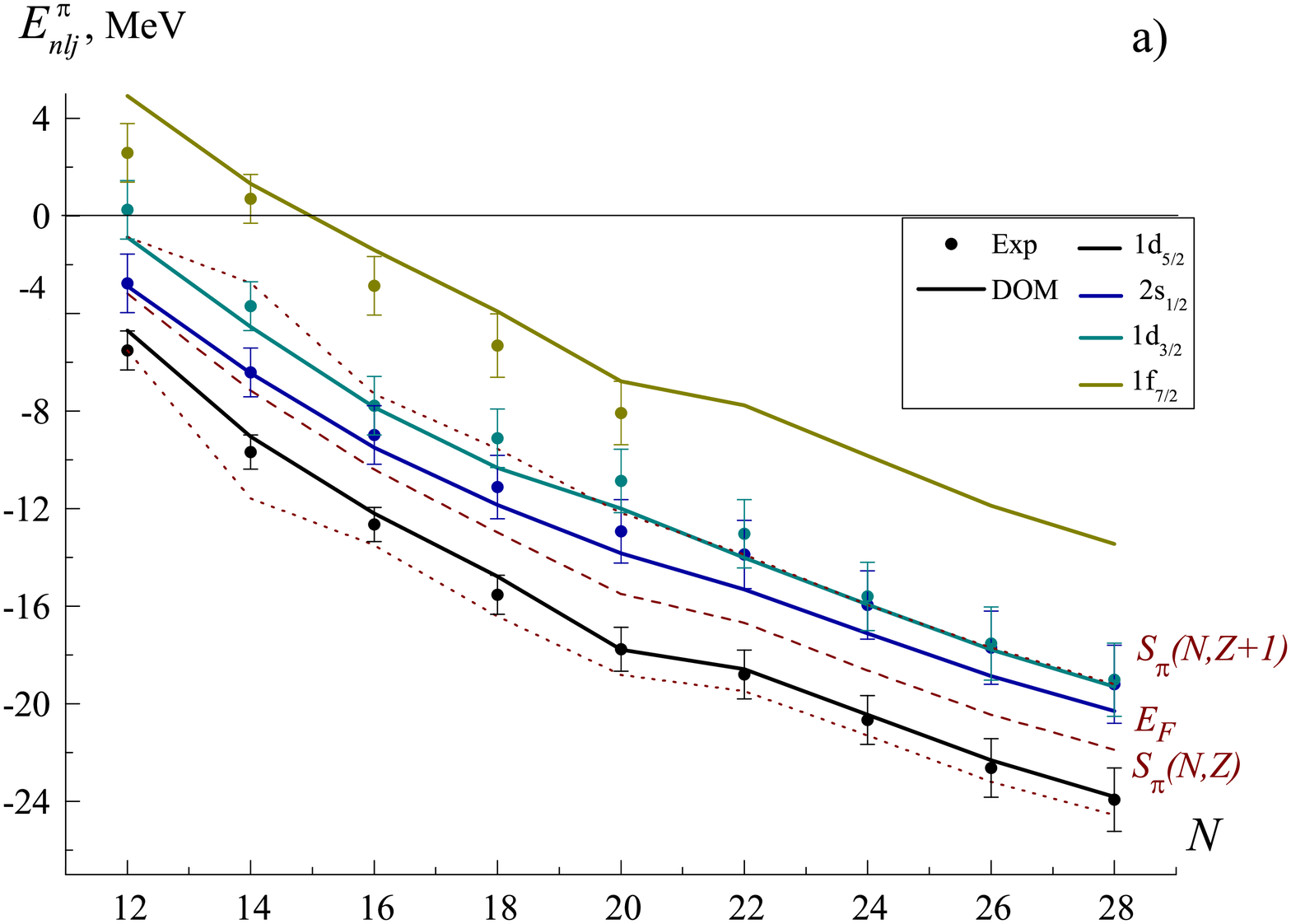}
}
\resizebox{8.5 cm}{!}{%
\includegraphics{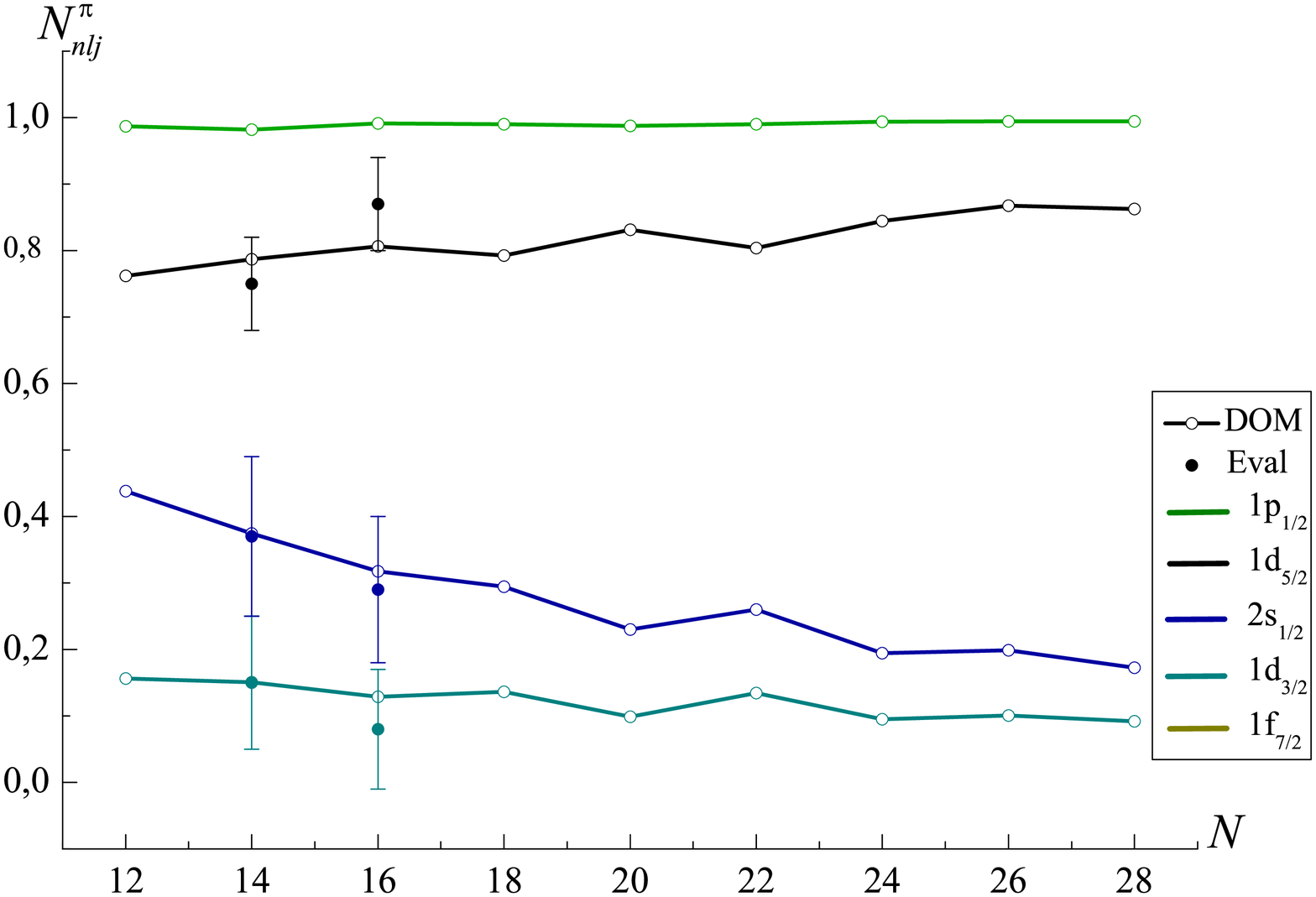}
}
\caption{(Colour online) Proton single particle energies $E^{\pi}_{nlj}$ (a) and occupation probabilities $N_{nlj}^{\pi}$(b) of Si isotopes. The error bars reflect the evaluation uncertainty.}
\label{fig:po}       
\end{figure}

Such a behavior of the evaluated energy $E_{nlj}^{eval}$ of the $1d_{3/2}$ state differs from that in Ca isotopes. According to experimental data \cite{BB05}, $1d_{3/2}$ state becomes more bound in Ca isotopes with $N$ increasing from 22 to~28. Global parameters of the imaginary potential \cite{KD03} lead to the analogous deepening of this state in Si isotopes as well. Agreement with the evaluated energies is improved if one assumes that the absorption increases in the isotopes of Si with $N$ increasing at $N > 20$ in contrast with the global parameters \cite{KD03}. The depth parameter $W_d$ \cite{KD03} decreases for neutrons as the neutron-proton asymmetry $(N-Z)/A$ grows in accordance with the Lane potential \cite{L62}. In nuclei near the neutron drip line, it is reasonable to expect  an increase of the density of nuclear levels
near the Fermi energy and of the absorption consequently. The dependence of the surface absorption $W_d$ in nuclei far from the beta-stability valley is discussed in \cite{CM07}. The neutron absorption $W_d$ was proposed to be independent of neutron-proton asymmetry in nuclei at $N > Z$ in order to describe neutron experimental spectroscopic factors of Ca isotopes. According to the proposal, we  fixed the parameter $d_1 = 16$~MeV equal to the one for isotope $^{28}$Si with $N=Z$ \cite{KD03}. As a result, the energies of $1d$ and $2s$ states of $^{38}$Si isotope, for example, increased by about 0.8~MeV and improvement of the agreement with the evaluated energies was obtained. However, it was not enough to achieve the agreement within the accuracy of evaluated data. For this, parameters $d_3$ and $w_2$ were decreased from 11.5 and 76.6~MeV \cite{KD03} to 7 and 40~MeV respectively. As a result, the maximum of dependence $W_d(E)$ increased sufficiently from 5.7~MeV \cite{KD03}  to 10.7~MeV and the energies  $-6.30$  and $-7.99$~MeV  of  $1d_{3/2}$ and $2s_{1/2}$ states respectively were obtained for $^{38}$Si. These values agree with the evaluated data within its accuracy and leads to agreement with the evaluated value of $fd$ splitting $\Delta_{\nu}(1f_{7/2}-1d_{3/2})=3.0$~MeV. It should also be noted that an increase in the radius parameter $r_{HF}$ also leads to a rise of the levels $1d_{3/2}$ and $2s_{1/2}$.

\begin{figure}
\resizebox{9 cm}{!}{%
\includegraphics{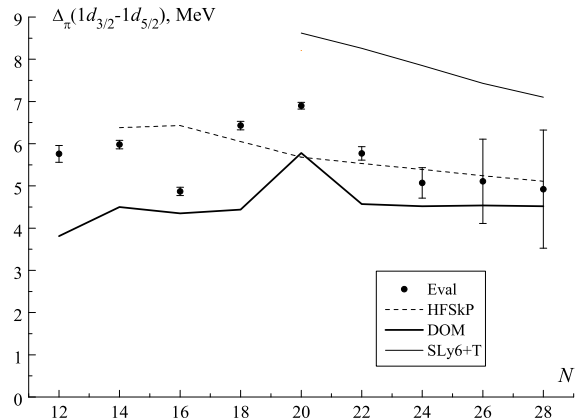}
}
\caption{Splitting between proton single-particle energies $1d_{3/2}-1d_{5/2}$, thin solid line present the calculations from \cite{Tarpanov}. }
\label{fig:dd_p}      
\end{figure}

}

Evolution of the proton single particle energies $E_{nlj}$ and the occupation probabilities $N_{nlj}$ is shown in Fig.~\ref{fig:po}. The calculated energies agree with the evaluated data except for the energies of the unoccupied $1f_{7/2}$ state which are slightly overestimated compared to $E_{nlj}^{eval}$.

Let us pay particular attention to the evolution of the proton $2s_{1/2}$ state. Both evaluated and calculated energies of this state are close to Fermi energy in $^{26}$Si. $2s_{1/2}$ level moves upward from $E_F$ with  increase of $N$ and approaches $-S_p(N,Z+1)$. This leads to the decrease of the occupation probability of this state and formation of $Z=14$ magic number. The difference between $N_{nlj}$ of the $1d_{5/2}$ and  $2s_{1/2}$ states increases from 0.32 in $^{26}$Si to 0.60 in $^{34-42}$Si (see Fig.~\ref{fig:po}). It was shown in \cite{Mu11} that protons displayed much
stronger $W_d$ dependence on neutron-proton asymmetry  than neutrons. Assumption of the stronger dependence of $W_d$ on $(N-Z)/A$ in comparison with \cite{KD03} leads to reduction of the break between $N_{nlj}$ for $1d_{5/2}$ and $2s_{1/2}$ states and also $Z = 14$ energy gap as $N$ increases.
The structure of nuclei in vicinity of $^{42}$Si was studied experimentally in the nucleon knockout reaction from secondary beams of exotic nuclei \cite{FW05,SG14}. It was shown  that the proton subshell closure at $Z = 14$ is well-developed for such nuclei.

In Fig.~\ref{fig:dd_p}, the $dd$ splitting of proton spin-orbit partners $1d_{3/2}-1d_{5/2}$ is displayed. The evaluated $dd$ splitting  reduces weakly  with  increasing $N$ and the peak at magic number $N = 20$ is observed. Values of the proton $dd$-splitting of the present paper are about 6~MeV and substantially less than that of \cite{Cottle}, where the $dd$-splitting in $^{28}$Si exceeds 10 MeV. In \cite{Cottle} the splitting estimations were obtained from  Si$(t, \alpha)$ pick-up reaction data for $1d_{5/2}$ state \cite{Pe87} and Si$(^3$He$, d)$ stripping reaction data  for $1d_{3/2}$ state \cite{Dj83,Ve90} separately, while in our analysis about 35 experiments were considered jointly to get the evaluated data for these nuclei.

Of greatest interest is the change in splitting as  neutrons fill the $1f_{7/2}$ shell. Evaluated reduction of the proton $dd$ splitting from $^{34}$Si to $^{42}$Si is about 1.9~MeV. The reduction was shown \cite{Tarpanov} to be attributed to the effect of a tensor component in the effective interaction. Calculation results from \cite{Tarpanov} in non-relativistic  Skyrme-Hartree-Fock approach using SLy6 interaction with additional tensor component is shown by a thin solid line in Fig.~\ref{fig:dd_p}, the corresponding decrease of $dd$ splitting is 1.5~MeV. For comparison we perform calculations for the  Si isotopes chain within Hartree-Fock model with SkP parameter set (HFSkP) \cite{SkP} without the tensor interaction (dash line in Fig.~\ref{fig:dd_p}). The SkP parameter set fails to reproduce the deformed $^{28}$Si structure, but for neutron-rich isotopes with $A>32$ it gives the values of binding energies with accuracy of about 7\%. HFSkP calculations demonstrated smooth reduction of the $dd$ splitting of about 0.6~MeV. The  proton $dd$ splitting calculated by DOM is underestimated when compared with the evaluated data. It is almost constant with a peak at $N = 20$ which is a result of the parameter $E_0 \neq E_F$ selection. Thus reduction of the proton $dd$ splitting, was obtained by DOM taking into account the double magicity of $^{34}$Si in which (sub-) shell closure $N=20$ and $Z=14$ occurs. The DOM parameters from Tabl.~\ref{tab:3p} with the imaginary part \cite{KD03} lead to the 1.2~MeV reduction of the proton $dd$ splitting in $^{36}$Si in comparison with $^{34}$Si.  Peculiarities of the neutron-proton interaction in $^{34}$Si nucleus with closed proton subshell $Z=14$ and neutron shell $N=20$ leads to the increase of the energy range around the Fermi energy where imaginary part of DOM is zero. As a result, the last predominantly occupied proton state $1d_{5/2}$ becomes more bound whereas the first predominantly unoccupied states $1d_{3/2}$ and $2s_{1/2}$ becomes less bound. Thereby, the proton $dd$ splitting increases in $^{34}$Si compared to the neighbouring isotopes. Contrariwise, the excitation across $N = 28$ shell in $^{42}$Si should lead to the absence of substantial increase of the proton $dd$ spin-orbit splitting in this nucleus. The evaluated decrease of the splitting from $^{34}$Si to $^{42}$Si is consistent with the  damping of $N = 28$ shell closure in $^{42}$Si isotope near the neutron drip line. 

Another feature of proton single-particle evolution is degeneration of $2s_{1/2}$ and $1d_{3/2}$ states in $^{38,40,42}$Si, which is expressed in approach of the $2s_{1/2}$ level to the $1d_{3/2}$ level and thus leads to the widening of the particle-hole gap between $1d_{5/2}$ and $2s_{1/2}$ states. This result reflects the experimental data \cite{SG14} which supports the notion that $Z = 14$ subshell closure persists up to $^{40}$Si in spite of the enhanced collectivity in the region of $^{42}$Si. The near-degeneration was also obtained experimentally in $^{42}$Si region \cite{FW05}. The calculation by DOM shows that the gap between $2s_{1/2}$ and $1d_{3/2}$ states decreases (see Fig.~\ref{fig:po}, a) from 2.0~MeV in $^{26}$Si to 1.0~MeV in $^{42}$Si, but the complete degeneration was not achieved.

\begin{figure}
\resizebox{7 cm}{!}{%
\includegraphics{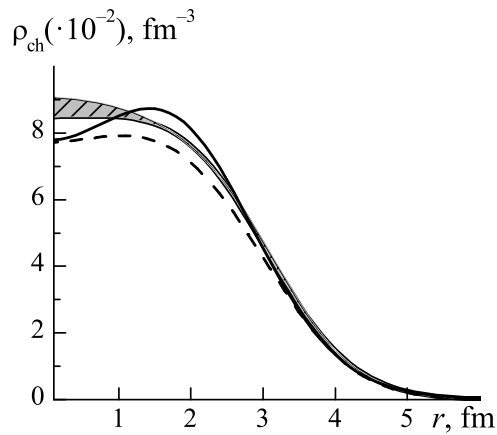}
}
\caption{Charge densities in $^{28}$Si. Results of DOM (solid line) and HFSkP calculations (dashed line) in  comparison with experimental results from \cite{DV87}. }
\label{fig:ch_d}      
\end{figure}

\subsection {Proton density distribution of Si isotopes}
\label{sec:4}

 To evaluate the DOM's \cite{MS91} ability to describe density distributions and rms radii of nuclei we calculated the charge density distribution $\rho_{ch}(r)$ of $^{28}$Si using $N_{nlj}$ (\ref{NDOM}) and normalized it to $Z=14$.  The distribution is shown in Fig.\ref{fig:ch_d} in comparison with the experimental data from \cite{DV87}  and results of  HFSkP computation. Some depletion of the central charge density was obtained. Corresponding rms charge radius 3.11~fm is in very good agreement with the experimental radius $r_{ch}$ = 3.1224 $\pm$ 0.0024~fm \cite{An04}. As it was mentioned above, the SkP parameters set fails to reproduce the deformed $^{28}$Si structure. It is possible origin ofobtained discrepancy between charge density distribution clalculated by HFSkP and the experimental data.
 
  It is known from electron scattering that the proton central density of stable nuclei is approximately constant 0.16~fm$^{-3}$ and almost independent of mass number $A$. Meanwhile, depletion of proton central density in $^{34}$Si is unexpected from this point of view but it was obtained confidently by both relativistic and non-relativistic models \cite{GG09} as well as state of art \textit{ab initio} calculations \cite{DS17}. Low population of the proton $2s_{1/2}$ state is the cause of the bubble-like structure of this nucleus. The depletion was recently investigated experimentally \cite{ML17} by one-proton removal reaction technique using unstable $^{34}$Si isotope secondary beam. It was obtained that $2s_{1/2}$ proton state was occupied with the probability of $0.17\pm 0.03$.  The values $N_{nlj} = 0.17$ and 0.23  calculated for $2s_{1/2}$ state using expressions (\ref{NDOM}) and (\ref{BCS}) accordingly agree with this experimental one. Also the experimental indication of the proton bubble-like structure was obtained in  \cite{BS14}. Addition of 2~protons leads to almost completely occupied $2s_{1/2}$ state in $^{36}$S and standard proton central density without any bubble-like structure. The strong reduction of neutron $2p_{3/2}-2p_{1/2}$ spin-orbit splitting was experimentally observed in $^{35}$Si nucleus in comparison with $^{37}$S. Because of the dependence of spin-orbit interaction on the density distribution, the reduction can be attributed just to the bubble-like structure of $^{34}$Si. 

\begin{figure}
\resizebox{8 cm}{!}{%
\includegraphics{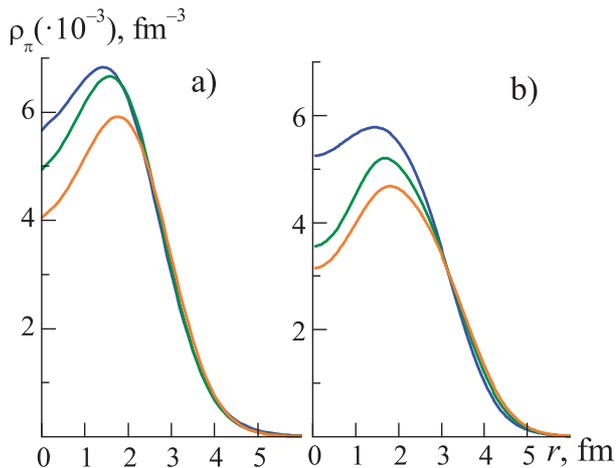}
}
\caption{Proton densities in $^{28, 34,42}$Si isotopes, DOM (a) and HFSkP (b) calculations. }
\label{fig:p_dns}      
\end{figure}

Experimental evidence of  $Z=14$ subshell closure up to $^{42}$Si \cite{FW05,SG14} allows one to expect bubble-like proton structure in Si isotopes with $A > 34$. The evolution of the proton density distribution of $^{28,34,42}$Si isotopes calculated by HFSkP and DOM with the $N_{nlj}$ (\ref{BCS}) is shown in Fig.~\ref{fig:p_dns}. The distributions are normalized to 1.  The proton central densities of $^{34,42}$Si isotopes, are significantly depleted so the bubble-like structure is formed. In comparison with $^{34}$Si, the maximum of density distribution of $^{42}$Si is slightly shifted to the region of larger radii. Additional neutrons pull the proton density from the central region towards the periphery, thus proton central density of $^{42}$Si is less than that of $^{34}$Si.

\section {Summary }
\label{sec:5}

Available experimental data on the neutron and proton stripping and pick-up reaction data on stable $^{28,30}$Si isotopes were analyzed jointly as well as the data on decay schemes of excited states of the unstable even isotopes $^{26,32-42}$Si. The neutron and proton single particle energies $E_{nlj}$ and occupation probabilities $N_{nlj}$ were evaluated for the states near Fermi energy. The results are compared with the calculations by mean field model with SkP effective interaction in HF approach and with the dispersive optical model potential.

The evolution of neutron single particle energies demonstrates the following features. In $^{30}$Si, the energies of $2s_{1/2}$ and $1d_{3/2}$ states are close to the neutron separation energies from $^{30}$Si and $^{31}$Si respectively indicating the new $N = 16$ magic number. The neutron $fd$ splitting proves the $N = 20$ shell closure in $^{34}$Si. In $^{42}$Si, the particle-hole gap calculated by DOM between $1f_{7/2}$ and $2p_{3/2}$ neutron states  decreased to 1.5~MeV in the comparison with the experimental gap of $5.5\pm1.1$~MeV in $^{48}$Ca. This agrees with the experimental evidence \cite{SG14} of excitations across $N = 28$ gap in $^{42}$Si. Comparison with the experimental reduction of about 330~keV per pair of protons removed from $^{49}$Ca \cite{Gau06} gave us grounds to assume that the gap reduces unevenly with $Z$ decreasing.

The evaluated  probabilities $N_{nlj}^{eval}$ of the proton $1d-2s$ states  in $^{28,30}$Si indicate the forming of $Z =14$ subshell closure  as $N$ increases. The distance between the calculated occupation probabilities $N_{nlj}$  of $1d_{5/2}$ and $2s_{1/2}$ proton states also increases from 0.3 in $^{26}$Si to 0.7 in $^{42}$Si and agrees  with the persistence of $Z = 14$ subshell closure with  increase of $N$. The evaluated data on single-particle energy evolution display the degeneration of $2s_{1/2}$ and $1d_{3/2}$ proton states which leads to the widening of the particle-hole gap between $1d_{5/2}$ and $2s_{1/2}$ states in $^{38,40,42}$Si. This result reflects the experimental data \cite{SG14} which supports the notion that $Z = 14$ subshell closure persists up to $^{40}$Si in spite of the enhanced collectivity in the region of $^{42}$Si. As a consequence, it is reasonable to expect central proton depletion due to the low occupation of the proton $2s_{1/2}$  state not only in $^{34}$Si but in Si isotopic chain with $N > 20$. In the present paper, the example of such depletion in $^{42}$Si was calculated by DOM and HFSkP.

Evaluated and calculated by DOM proton $dd$-splitting indicates that $N=20$ is a magic number. Values of the proton $dd$-splitting of the present paper are substantially less than that of \cite{Cottle}, in particular for $^{28,30}$Si. This result is important for the investigation of the spin-orbit and tensor parts of nucleon-nucleon interaction.

The results of calculations demonstrate that DOM is applicable for the calculations of evolution of single-particle nuclear characteristics over a wide range of $N$ and $Z$ numbers. Very good agreement of the charge radius of $^{28}$Si with the experimental data was achieved by DOM. So, the initial version of DOM \cite{MS91} can be successfully used to calculate rms radii of nuclei. The most substantial discrepancy of the calculated results from the evaluated data is manifested in evolution of the neutron energies of $1d_{3/2}$ and $2s_{1/2}$ states for $N > 20$. The missing of the experimental knowledge about fragmentation of these states may be one of the causes of the discrepancy. Another cause is the extrapolation of the DOP parameters found for stable nuclei on the region of nuclei far of beta-stability line. For nuclei with essential neutron excess, it may be necessary to correct the parameters of DOM, in particular, the imaginary part of the potential in comparison with the global parameters  found for stable nuclei. 

The authors are grateful to Prof. B.S.~Ishkhanov and Prof. D.O.~ Eremenko for their attention to this work. The authors also thank I.N.~ Boboshin, N.G.~Goncharova, D.E.~Lanskoy for useful discussions and suggestions. T.T. thanks S.V.~Sidorov for his help in manuscript preparing.

%
%
%
%

\end{document}